\documentclass[twocolumn]{aastex61}
\usepackage{color}
\usepackage{latexsym, graphicx, amssymb, longtable,epsf,ulem}

\newcommand{\Xie}{\color{black}}
\newcommand{\xie}{\color{black}}
\newcommand{\xxie}{\color{black}}

\newcommand\aastex{AAS\TeX}

\shorttitle{\aastex\ sample article}
\shortauthors{Zhang et al.}

\begin{document}

\title{Planet Formation in Highly Inclined Binary Systems. II.  \\{\Xie Orbital  Alignment {\xxie or Anti-Alignment} and Planet Growth Boost in Intermediate separation binaries}}

\author{Yapeng Zhang}
\affil{School of Astronomy and Space Science \& Key Laboratory of Modern Astronomy and Astrophysics in Ministry of Education, Nanjing University, 210093, China}
\author{Qingqin Li}
\affil{School of Astronomy and Space Science \& Key Laboratory of Modern Astronomy and Astrophysics in Ministry of Education, Nanjing University, 210093, China}
\author{Ji-Wei Xie}
\affil{School of Astronomy and Space Science \& Key Laboratory of Modern Astronomy and Astrophysics in Ministry of Education, Nanjing University, 210093, China}
\affil{To whom correspondence should be addressed. E-mail:  jwxie@nju.edu.cn}
\author{Ji-Lin Zhou}
\affil{School of Astronomy and Space Science \& Key Laboratory of Modern Astronomy and Astrophysics in Ministry of Education, Nanjing University, 210093, China}
\author{Hui-Gen Liu}
\affil{School of Astronomy and Space Science \& Key Laboratory of Modern Astronomy and Astrophysics in Ministry of Education, Nanjing University, 210093, China}
\author{Hui Zhang}
\affil{School of Astronomy and Space Science \& Key Laboratory of Modern Astronomy and Astrophysics in Ministry of Education, Nanjing University, 210093, China}

\begin{abstract}
Stars are commonly formed in binary systems, which provide a natural laboratory for studying planet formation in extreme conditions. 
In our first paper (Paper I) of a series \citep{Xie11}, we have shown that the intermediate stage $-$ from planetesimals to planetary embryos/cores $-$ of planet formation can proceed even in highly inclined binaries.  
Following Paper I, here we numerically study the late stage of terrestrial planet formation, i.e., from embryos to full planets, in binary systems of various orbital configurations. 
We identify an orbital alignment {\xxie or anti-alignment} effect; namely, although an inclined binary generally misaligns the planetary orbits with respect to the spin axis of the primary host star (i.e., causing large obliquity), it could align {\xxie or anti-align} the planetary orbits with respect to the binary orbit. 
{\xie Such an orbital {\xxie (anti-)} alignment effect is caused by the combination of orbital differential precession and self-damping, and }it is mostly significant in cases of intermediate binary separations, i.e., $a_B\sim40-200$ AU for {\xxie terrestrial} planet formation around 1 AU from the primary stars.  
{\Xie In such intermediate separation binaries, somewhat contrary to intuition,  the binary companion can aid planet growth by having increased the rate of collisions, forming significantly more massive but fewer planets. In the other two ends, the companion is ether too close thus plays a violently disruptive role or too wide to have significant effect on planet formation.
Future observations, which can discover more planet-bearing binary star systems and constrain their masses and 3-D orbital motions will test our numerical findings.}

\end{abstract}

\keywords{method: numerical -- planet and satellites: formation}

\section{Introduction} \label{sec:intro}
Hitherto, over 3500 exoplanets have been discovered and they are found to be very common around stars \citep{May11,  How12}. Combining with another well established view, i.e., most stars were born in binaries \citep{DM91, Rag10}, it is clear that one should be always aware of potential binary effects on planet formation. In fact, studying planet formation in binaries is crucial as it provides various extreme conditions for testing planet formation models \citep{Hag10, TH14}.

Planets in binaries are generally found in two configurations. One is called P type, where all the planets orbit two host stars (i.e., the host itself is a close binary, e.g., Kepler 16 \citep{Doy11}) . The other is called S type, where all the planets orbit one of the binary stars and the other star orbits the planets-host system as a companion, e.g., $\gamma$ Cephei AB \citep{Hat03}. Over 100 planets have been confirmed in binary star systems (see the catalog of planets in binaries\footnote {http://www.univie.ac.at/adg/schwarz/bincat\_binary\_star.html}), and most of them are classified as S type, which is the focus of this paper.

For planet formation in binary systems, one crucial parameter is the binary separation \citep{DB07}.  Close binaries (orbital semimajor axis $a_B<30-40$ AU)  can truncate the protoplanetary disk and severely reduce the mass of the planet forming materials \citep{AL94}. Furthermore, binary perturbations can pump up the orbital eccentricities of the planetesimals in the disk, which cause large collision velocities and thus inhibit the planetesimals from growing into planetary embryos \citep{The06}. Although observations have revealed evidence of planets being suppressed in close binaries \citep{Wan14, Kra16}, there are still a number of planets found in close binary systems with $a_B\sim20$ AU  (e.g., $\gamma$ Cephei AB \citep{Hat03}, $\alpha$ Cen AB \citep{Dum12}). Many works have been stimulated to understand how planets were formed in such close binaries, especially the intermediate process, i.e., from planetesimals to planetary embryos \citep{Hag06, HR07, KN08, The08, The09, The11, Xie08, Xie09, Xie10a, Zso11, RS15a}.

{\xie  On the other hand, for binaries with larger separations,  one does not expect there to be significant alignment between the proto-planetary disk and binary orbital plane if $a_B>30-40$ AU \citep{Hal94, Jen04, Mon06, Mon07}.  
The evolution of an inclined gaseous disk in a binary system is uncertain due to various uncertainties both in the physics of disk-star interaction and in the parameters of disk properties. 
In some conditions, planetary disk could be warped or even disrupted \citep{Lar96, FN10, Fra11}. 
In some conditions, the disk could undergo global damped Kozai–Lidov oscillations \citep{Mar14, Fu15}.  
Dissipations (e.g., viscous dissipation) could damp the disk inclination towards disk-binary alignment \citep{LO00, Bat00}.
Recently, \citet{ZL17b} found that such an alignment is effective for sufficiently cold disks (small scale height) with strong external torques but ineffective for the majority of star-disk-binary systems. 
Indeed, many misaligned binary disks, e.g, HK Tau \citep{JA14} have been found in recent years \citep{Wil14, Bri16, Fer17, Lee17}.  

Adopting a highly misaligned binary-disk configuration as the initial condition of planet formation, \citet{Mar09} and \citet{Xie11} (Paper I hereafter)  investigated the intermediate process of planet formation, i.e., from planetesimal to planetary embryos. }
As found in Paper I, planetesimals could jump inward and pile up within  a few AU from the primary star. In such an inner dense region, the perturbations from the binary stars are largely compensated by the damping of gas in the disk, thus providing conditions that are favorable for planetesimals growing up into planetary embryos.
In this paper, following paper I, we numerically investigate how these embryos could further grow up to full planets and how their final architectures depend on binary orbits. We note a previous study by \citet{Qui02} investigated the process from embryos to full planets in highly inclined binaries but with the binary separation being fixed at $\sim$ 20 AU as the Alpha Centauri AB system.  Another previous work by \citet{Qui07} investigated the embryo growing process in binaries of various separations, but it was restricted in coplanar cases.  
{\xie 
More recently, during the revision of this paper, \citet{ZL17a} investigated how the formation a gas giant planet could affect the final planetary orbital configuration in an  initially inclined binary system. 
}
Here, we extend  \citet{Qui02}  to a larger range of binary separations.    As we show below,  the binary separation is crucial, and various binary separations lead to diverse planetary architectures.  Of particular interest is that we identify an orbital alignment effect in intermediately separated  binaries.

This paper is organized as follows. In section \ref{sec:method}, we describe our simulation method, including the initial set-up of the embryos and the binary stars.  The results are presented in section \ref{sec:results}. The implications of the results are discussed and summarized in section \ref{sec:disc}.
  
\begin{figure}[ht!]
\includegraphics[width=0.47\textwidth]{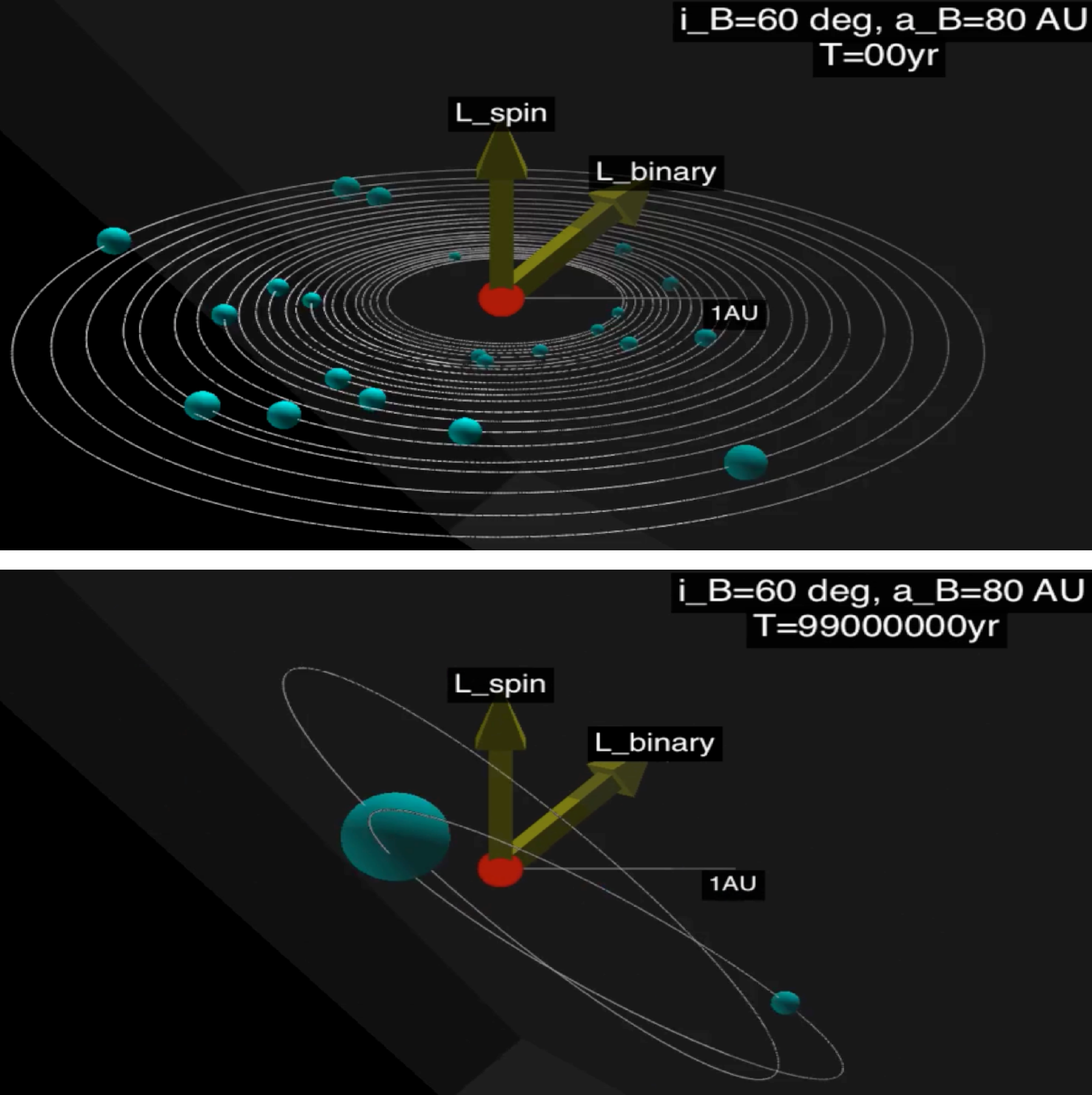}
\caption{Initial setup (top) and final state (bottom) of a typical simulation with $a_B=80$ AU, $i_B=60^\circ$. The orbits of planetary embryos (denoted with the cyan spheres) were initially aligned with respect to the spin of the primary (denoted with the red sphere) but finally aligned with respect to the binary orbit (not shown). The green arrows denote the normal lines of the primary spin plane ($L_{spin}$) and of the binary orbital plane ($L_{binary}$), respectively. These plots are two snapshots of an animation, which is available in the electronic version of the paper. \label{fig:sketch}}
\end{figure}
  
\begin{figure*}[ht!]
\includegraphics[width=\textwidth, height=0.82\textwidth]{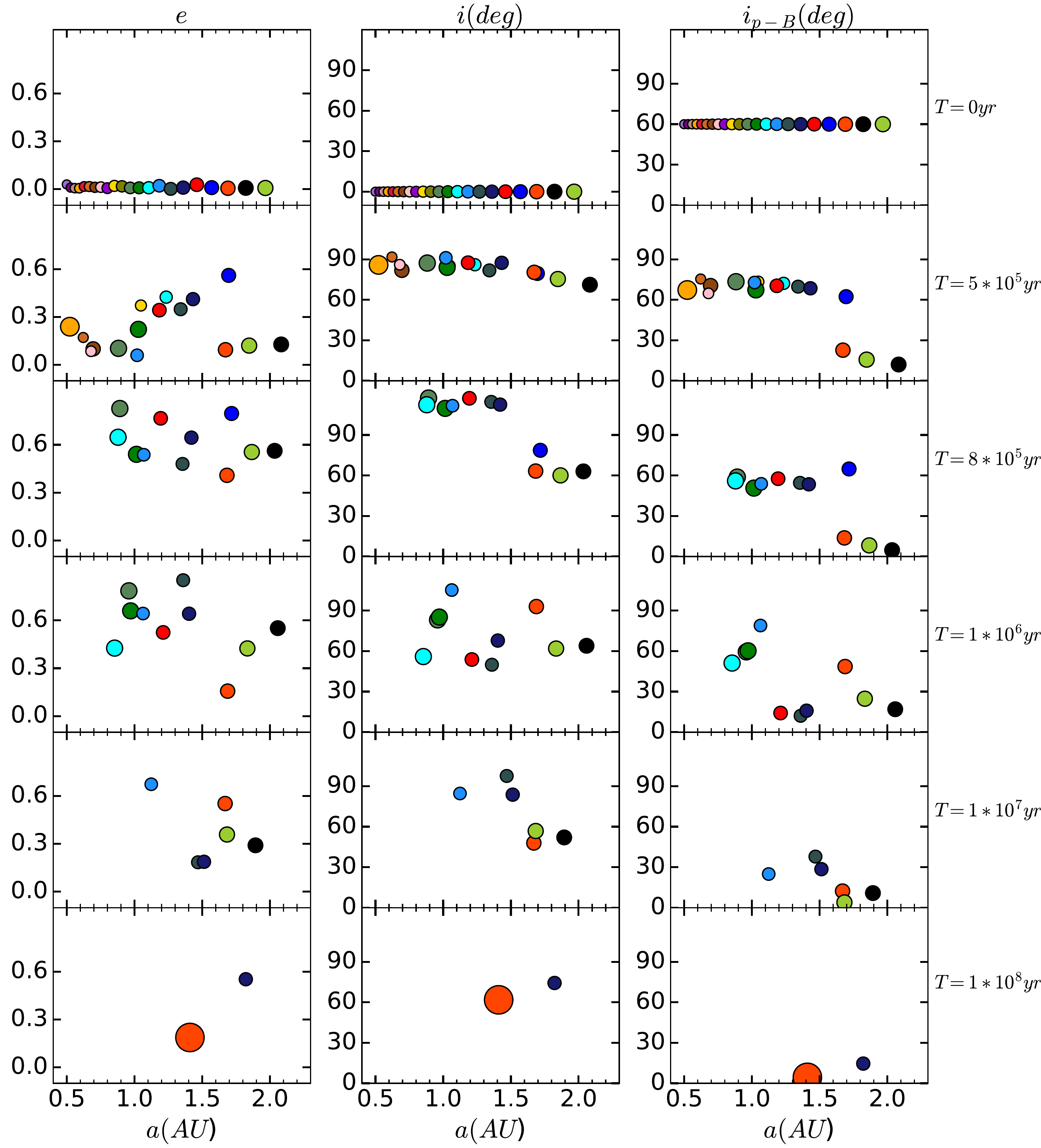}
\caption{Snapshots of the embryos for a typical case of $a_B=80$ AU, $i_B=60^\circ$ at different time $T$(yr), illustrating the growth of embryos and evolution of their orbital elements, including orbital eccentricities $e$, orbital inclinations ($i$) with respect to the initial protoplanetary disk plane and inclinations ($i_{p-B}$) relative to the binary orbital plane.  Note the sizes of the circles are scaled to the masses of the embryos and the colors are used to distinguish different embryos, which are corresponding to the colors in Figure \ref{fig:track}. \label{fig:snap}}
\end{figure*}

\begin{figure*}[ht!]
\includegraphics[width=0.99\textwidth]{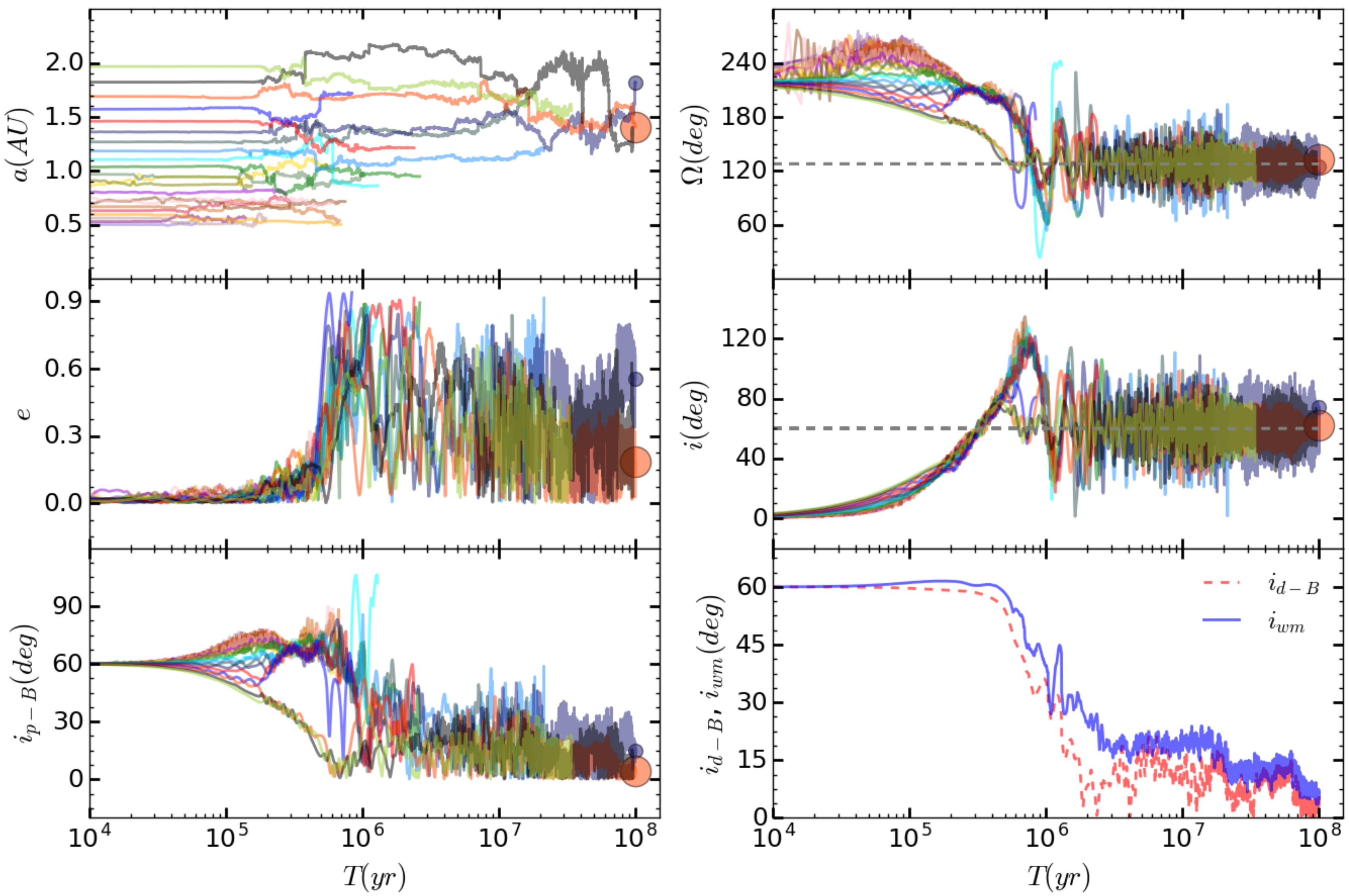}
\caption{Tracks of embryos for a typical case of $a_B=80$ AU, $i_B=60^\circ$, showing the evolutions of the planet masses ($M_{p}$) and orbital elements, including the semi-major axes ($a$), orbital eccentricities ($e$), longitude of ascending nodes ($\Omega$), inclinations ($i$) relative to the initial proto-planetary disk plane, inclinations ($i_{p-B}$) relative to the binary orbital plane, $i_{wm}$ (mass-weighted of $i_{p-B}$) and $i_{d-B}$ (inclination between the planetary disk as a whole and the binary orbital plane, i.e., the angle between the disk angular momentum and the binary orbital angular monentum). This is the same simulation case as plotted in Figure \ref{fig:snap} and the colors used here are corresponding to the circles in Figure \ref{fig:snap}. The orange and blue circles (see also in Figure \ref{fig:snap}) highlight the two planets formed at the end of simulation. The dashed horizontal lines denote the orbital inclination ($i_B=60^\circ$) and longitude of ascending node ($\Omega_B\sim120^\circ$) of the binary orbit with respect to the initial proto-planetary disk. As can be seen, the planets' orbits tend to be aligned with the binary orbit, i.e., $i\sim i_B$ and $\Omega \sim \Omega_B$ or $i_{p-B}\sim0$ and $i_{d-B}\sim0$. \label{fig:track}}. 
\end{figure*}

\begin{figure}
\includegraphics[width=0.47\textwidth, height=0.45\textwidth]{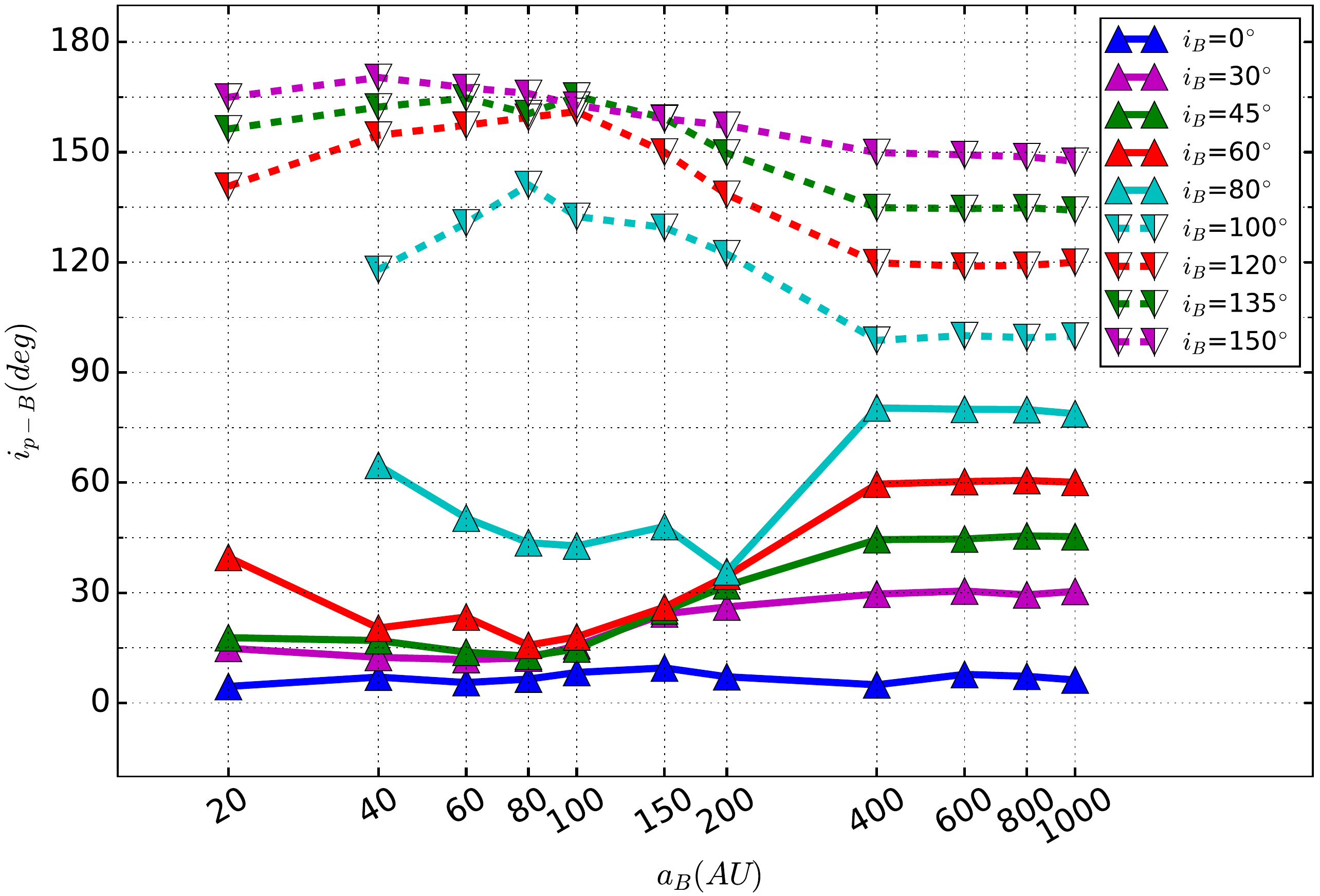}
\caption{Orbital {\xxie (anti-)}alignment effect, quantified by the statistic $i_{p-B}$, and its dependence on binary orbital parameters, i.e., $a_B$ and $i_B$.  Each  $i_{p-B}$ shown here is the average final orbital inclination (with respect to the binary orbital plane) of 10 simulations for each pair of $a_B$ and $i_B$. {\Xie The orbital {\xxie (anti-)}alignment effect is most significant for intermediate separation binaries ($a_B\sim$40-200 AU).  For wider binaries ($a_B\ge$200 AU), the binary perturbation is weaker and the self-gravity keep the planetary embryos as a rigid disk precessing with an invariable $i_{p-B}$}.    \label{fig:depend}}
\end{figure}

\section{Method} \label{sec:method}

In this paper, we numerically investigate the late stage (i.e., from embryos to full planets) of terrestrial planet formation in the S type binary systems.
According to the binary survey of the solar neighborhood\citep{DM91, Rag10}, we set typical binary configurations with the primary star having a mass of solar mass, i.e., $M_A=1 M_\odot$, the binary mass ratio $M_B/M_A=0.4$ and the orbit eccentricity being fixed as $e_B = 0.4$.  We vary the binary semimajor axis ( $a_B \sim 20-1000$ AU) and the binary orbital inclination ($i_B\sim0-180^\circ$ with respect to the initial plane of the planetary embryo disk) to investigate their effects on the final architecture of the planetary system.  

For the initial protoplanetary disk, following \citet{Kok06}, we assume a power law solid surface density profile, i.e.,
\begin{eqnarray}
{\Sigma} = {\Sigma}_1 {(\frac{a}{1AU})}^{-3/2} g/cm^{2},
\end{eqnarray}
where ${\Sigma}_1$ is the reference surface density at 1AU.
We further assume a chain of planetary embryos have accreted all the solid materials in the disk and reached their isolation masses \citep{Kok06},
\begin{eqnarray}
M_{iso} \simeq 0.16(\frac{b}{10})^{3/2}(\frac{\Sigma_1}{10})^{3/2}(\frac{a}{1AU})^{(3/4)} M_{\oplus},
\end{eqnarray}
where $b$ measures the mutual orbital separations of these embryos in their mutual Hill radius and  $ M_{\oplus}$ is the Earth mass. In this paper, we adopt $b=10$ and  ${\Sigma}_1=10$, representing an initial disk similar to the Minimum Mass of Solar Nebula (MMSN), which ends up with 22 embryos  between 0.5 and 2.0 AU from the primary star. The total mass of these embryos is $M_{tot}\sim3.6 M_{\oplus}$.
In our simulations, the bulk densities of embryos are $\rho=3g/cm^{3}$, and the initial eccentricities ($e$) and inclinations ($i$) of embryos are given by the Rayleigh distribution with dispersion $ <e^2>^{1/2}=2<i^2>^{1/2}=0.01(\Sigma_1/10)^{1/2}$ \citep{Kok06}. The remaining  angular orbital elements of all bodies are randomly generated in a range from 0 to 360$^\circ$.
 
Note that, above initial conditions for planetary embryos are \emph{only} well justified in single star systems \citep{Kok06}. In fact, the initial condition in binary system is poorly known and it is out of the scope of this paper. Nevertheless, setting the initial condition the same as in single systems would make it straightforward to compare the results of different simulations and  identify the effects of binary stars (as compared to single stars) on planet formation.
 
We performed simulations using the N-body code MERCURY \citep{Cha99}. 
In the majority of our simulations, we adopted the WB {\xie (Wide Binary)} algorithm, a modified symplectic integrator, intended for close encounters among embryo bodies in the S type binary systems \citep{Cha02}. For comparison, the simulations without the presence of the stellar companion are also performed, in which the HYBRID algorithm was used.
The duration of integration is set up to 100 million years ($\approx 3 \times 10^8$ period of the innermost embryo orbit $P_{in}$), and the time step is 7 days. 
Due to the chaotic nature of N-body simulation, for each pair of parameters ($a_B$ and $i_B$), we performed 10 simulations with other angular orbital elements randomly drawn in a range from 0 to 360$^\circ$ to access the statistics of the results. {\Xie For the specific case shown in Figures 2 and 3, the argument of pericenter, the longitude of the ascending node and the mean anomaly are 354$^\circ$, 128$^\circ$ and 26$^\circ$ respecttively. }  Figure \ref{fig:sketch} shows two   snapshots of an animation, illustrating the initial setup and final state of a typical simulation. 
{\xie In addition, we also performed a set of simulation using the Bulirsch-Stoer (BS) algorithm to cross-check the results (See appendix).}

 {\Xie In all of our simulations, we ignored the effect of general relativity (GR), which could cause orbital precession and potentially suppress the dynamical pumping, e.g., binary Kozai effect. For the typical planet formation site at $a_p=1$ AU as considered in this paper, the GR precession timescale is $\sim 3.3\times10^7$ yr corresponding to the Kozai timescale at $a_B\sim300$ AU. Therefore, we expect that GR will suppress the binary secular perturbation for wide binaries but will not have large impact on the orbital alignment effect as shown below, which is most significant in cases of intermediate binary separations, i.e., aB=40-200 AU.}

\section{Results} \label{sec:results}

\subsection{An Orbital Alignment Effect} 
{\xie We take the case of $a_B=80$ AU, $i_B=60^\circ$ as an example (same as in Fig.\ref{fig:sketch}) and plot Figure \ref{fig:snap} and Figure \ref{fig:track} to have a scrutinous inspection of the simulation. Figure \ref{fig:snap} shows the snapshots of embryos in the distributions of orbital eccentricities ($e$), inclinations ($i$, with respect to the initial orbital plane, i.e., the equator plane of the primary star) and inclinations relative to the binary orbital plane ($i_{p-B}$). Figure \ref{fig:track} shows the evolutions of $a$ (semi-major axis), $e$,  $i_{p-B}$, $i$,  $\Omega$ (longitude of orbital ascending node), $i_{\rm wm}$ (mass weighted mean of $i_{p-B}$) and $i_{d-B}$ (inclination between the total planetary disk angular momentum and the binary orbital angular momentum).  }

As can be seen in Figure \ref{fig:snap} and Figure \ref{fig:track}, all the planetary embryos were initially in the protoplanetary disk with nearly circular and coplanar orbits. If we further assume that the rotation of the primary star was initially aligned with the protoplanetary disk, then the obliquities of embryos were equal to their orbital inclinations.   As the system evolved, the orbits of embryos became chaotic under the perturbations of the binary star and their self-gravity. They  were pumped to orbits with large eccentricities and inclinations.  The orbital-crossing embryos collided and grew bigger.  At the end of simulation, two planets were formed. Interestingly,  although the orbits of the two final planets became misaligned with respect to the initial orbital plane (thus large obliquities), they were nearly aligned with respect to the binary orbital plane. In the follows, we present our explanation to physically understand such an orbital alignment effect.  

{\xie
The above orbital evolution is composed of two key processes, i.e., a dynamically pumping process and a dynamically damping process.  
In the dynamically pumping process, embryos' orbital eccentricities were pumped up due to the combination of binary perturbations, e.g., via Kozai mechanism \citep{WM03, FT07} and their self-gravitational perturbations. 
Moreover, the binary secular perturbation caused orbital differential precession (due to differential precession rates of embryos with different orbital semi-major axes) of the planetary embryos, which randomized their orbital orientations. 
A direct outcome of  orbital differential precession is that the planetary disk as a whole will be on average aligned with  the binary orbital plane, though individual planetary embryos still being highly misaligned.   
This dynamically pumping process can be mostly seen in the first two million years of the dynamical evolution as shown in Figure \ref{fig:track}.  
At $t\sim2\times10^6$ yr, embryos were pumped up to large eccentricities ($e$) and their orbits were randomized with a large range of inclinations ($i$) and orientations ($\Omega$), leading to alignment between the planetary disk and the binary orbit ($i_{d-B}\sim0$).

 The dynamically pumping process was companioned by a dynamically damping process. 
Once the eccentricities were pumped up, orbital crossing then enhanced the rates of close encounters, which would damp the orbital inclinations via the so-called dynamical friction effect \citep{Obr06} or direct collisions. 
This damping effect can be clearly seen in the bottom right panel of Figure \ref{fig:track}; the mass weighted inclination, $i_{wm}$,  reduced from $\sim60^\circ$ to $\sim5^\circ$.   
Note the damping process did not directly damp $i_{p-B}$ or $i_{wm}$ but the mutual orbital inclinations of embryos. 
In order to damp $i_{p-B}$ or $i_{wm}$, the mid-plane of the embryo disk should be aligned with the binary orbital plane, which was realized by the orbital differential precession in the dynamically pumping process. It worth noting that collisions were roughly treated as perfectly inelastic in all the simulations. In reality, relative velocities of embryos would be rather high due to binary perturbations, and real collisions would cause lots of fragments. Nevertheless, including fragments would be likely to enhance orbital damping thus reinforce the orbital alignment via dynamical friction. 

In the appendix, we performed more simulations (Fig. \ref{fig:a1}--\ref{fig:a4}) by isolating various factors, i.e., collision and self-gravity, which separated the pumping and damping processes and helped us understand the dynamics of the orbital alignment effect.  
}  

\begin{figure}[ht!]
\includegraphics[width=0.47\textwidth, height=0.85\textwidth]{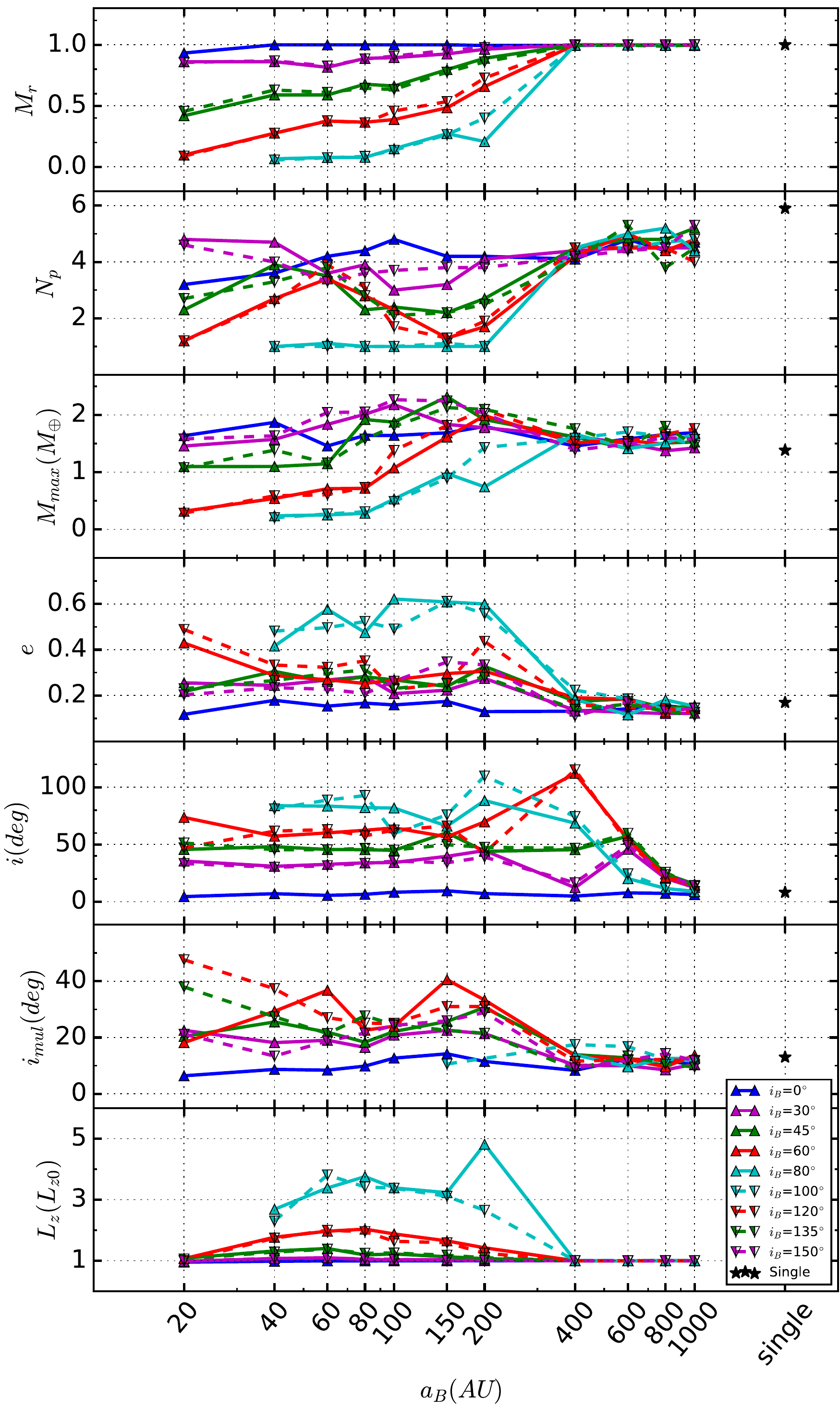}
\caption{Similar to Figure \ref{fig:depend}, but showing other statistics and their dependences on binary orbital parameters, i.e., $a_B$ and $i_B$.  The definitions and values of these statistics are presented in Table \ref{tab:A}.   \label{fig:other}}
\end{figure}

\subsection{Parameter Dependencies} 
We find that the above orbital alignment effect depends on the binary orbital parameters.  In Figure \ref{fig:depend}, we plot the mean orbital inclination ($i_{p-B}$) between the finally formed planets and the binary star for different initial $a_B$ and $i_B$.  We see that distributions of the final $i_{p-B}$ are approximately symmetrical with respect to $i_B=90^\circ$, which is expected from the theory of secular dynamics. As long as the initial binary orbit is not extremely inclined ($i_B\sim80^\circ$--$100^\circ$), the orbital alignment (actually it is anti-alignment if $i_B>90^\circ$) effect is mostly significant in binaries with intermediate separations ($a_B\sim40 $--$200$ AU).   The degree of (anti-)alignment is nearly independent of $i_B$ and achieves to an deviation as small as $\sim10^\circ$--$20^\circ$ from the binary orbital plane. 

Binaries with extremely inclined orbits ($i_B\sim80^\circ$--$100^\circ$) or relatively small separations (e.g., $a_B<40$ AU), induced too strong perturbations, under which most planetary embryos were over-pumped to extremely eccentric orbits and lost by colliding with the central star. {\xie Thus the damping via dynamical friction and collision was inefficient afterwards and the orbital alignment effect is not significant as shown in Figure.~\ref{fig:depend}}.  At the other end, for binaries with very wide separations ($a_B>200$ AU), the secular perturbation from the binary stars is weaker than the secular perturbations induced by the mutual gravity of embryos. In such a case, orbital differential precession via secular perturbation was largely quenched, and the embryos maintained the initial inclination relative to the binary (i.e., $i_{p-B}=i_B$ as in Fig.~\ref{fig:depend}) and precessed as a rigid disk \citep{Tak08}. 

In the appendix, we performed more simulations (Fig. \ref{fig:a5} and Fig. \ref{fig:a6}) to show how the orbital alignment effect is not significant in very close/wide binaries.

\subsection{{\Xie Other Effects: e.g., Binary Aids Planet Growth}}
Besides the orbital alignment effect, we can obtain more insights into how binary stars affect planet formation by studying more statistics. These statistics are listed in Table \ref{tab:A} and plotted in Figure \ref{fig:other}.
For comparison, we also performed a set of ``single" simulations, in which the companion star was removed and only a single star left in the system as in our solar system. The results of the ``single" simulations are also shown in Figure \ref{fig:other}.  From bottom to top, we describe and analyze each panel in the follows. Cross-analyzing different panels improves our understanding of the results as a whole.

In the $L_z$--$a_B$ panel,  the Z-component of angular momentum  per unit mass, $L_z$,  generally conserves in the cases of close ($a_B<40$ AU) and wide ($a_B>200$ AU) binaries but  it is significantly enhanced in binaries with intermediate separations. This is in line with the orbital alignment effect shown in Figure \ref{fig:depend}. 

In the $i_{mul}$--$a_B$ and $e$--$a_B$ panels,  there is a general trend  that $i_{mul}$ {\Xie (average mutual orbital inclination of remaining planets)} and $e$ increase for binaries with more inclined and closer orbits. Nevertheless, there seems a plateau in the middle of the global trend, i.e., $i_{mul}$ and $e$ maintain moderate for intermediately separated binaries. This is in line with the efficient collisional damping of the orbital alignment effect.

In the $i$--$a_B$ panel, it shows $i\sim i_B$ for $a_B<200$ AU due to the orbital alignment effect. For wider binaries with $a_B>200$ AU, the orbital alignment effect is inefficient due to the weak secular perturbation from the binary, and thus the embryos precess as a rigid disk with $i$ oscillating in a range from 0 to 2$i_B$. Since the oscillation time scale increases rapidly with $a_B$,  most $i$ in wide binaries ($a_B>200$ AU ) are still in their low phases at the end of simulations.

In the $M_{max}$--$a_B$ panel, it shows that the maximum mass of a planet finally formed in binaries can be even larger than that in the single star system. This is most significant for those binaries with intermediate separations and moderate inclinations. For example, in a binary with $a_B=100$ AU and $i_B=30^\circ$, the mean maximum mass $M_{max}\sim 2.2 M_\oplus$, which is about 60\% higher than that in the single star system. The explanation is similar to the orbital alignment effect as shown in Figure \ref{fig:depend}, that moderate perturbations can enhance collision rate and thus boost planet formation.

In the $N_p$--$a_B$ panel, it shows that the numbers of planets formed in binaries are always smaller than that in the single star system. There are two reasons for this result. First, moderate binary perturbations can enhance collision, leading to fewer but bigger planets formed in the system.  This is clearly seen in binaries with intermediate separations and moderate inclinations.  The other reason is that binary perturbations cause significant embryo loss,  leading to fewer and smaller planets left in the system. This is dominant for binaries with very small separations and/or extremely high inclinations. 

In the $M_r$--$a_B$ panel, it shows that the remaining mass fraction at the end of simulation, $M_r$, is generally smaller for binaries with smaller separations and/or larger inclinations.  Systems with lower $M_r$ underwent more embryo-loss, and thus had less damping and growth by collisions.

\section{Summary and Discussion} \label{sec:disc}
In this paper, we investigate the late stage of terrestrial planet formation starting from a chain of planetary embryos in binary systems of various binary separations ($a_B$=20$-$1000 AU) and orbital inclinations ($i_B=0$$-$$180^\circ$).  We identify an orbital {\xxie (anti-)}alignment effect, namely a highly inclined configuration between embryos' orbits and the binary orbit could eventually evolve to be nearly coplanar. Such an orbital {\xxie (anti-)}alignment effect is mostly significant for binaries with intermediate separations. For terrestrial planet formation taking place around 1 AU as studied here, the intermediate binary separation corresponds to $a_B\sim$ 40$-$200 AU.  It is worthy noting that such an orbital {\xxie (anti-)}alignment effect is to {\xxie (anti-)}align the planetary orbital plane with respect to the binary orbital plane, which misaligns the planetary orbital plane with respect to the stellar spin axis. As shown in Figure \ref{fig:other},  such a by-product misalignment leads to an obliquity of $\Psi\sim i_B$.

{\xie
The orbital alignment effect identified in this paper resembles the orbital alignment as mentioned in the introduction \citep{LO00, Bat00, Fu15, ZL17b}, though they are not exactly the same. 
The latter operates for a gaseous disk and the former for a gas free disk.
Nevertheless, both alignments could be treated as the outcome of dissipation associated with disk warp/twist. 
In the latter case, the dissipation is mainly from gas viscosity and disk warp is due to the binary torque. 
While in the former case, the dissipation is from dynamical friction and collision, and the warp of embryo disk is due to the orbital differential precession caused by the binary secular perturbation.   
Systems which fails to be aligned during gas disk phase could still have chances to be aligned after the gas disk dissipates.
}

Stellar binaries have been invoked in some other mechanisms  to explained the large obliquities observed in exoplanets.  For examples, a highly inclined binary can lead to large planetary obliquities via Kozai mechanism \citep{WM03, FT07} or by tilting the protoplanetary disk \citep{Bat12}.  In those mechanisms, it was expected that the planetary orbits with large obliquities are misaligned with respect to the binary orbital plane.  However, our results suggest that such an expectation is not always necessary. At least in some circumstance as shown in this paper,  a highly inclined binary, which misaligned the planetary orbits with respect to the primary rotation axis (i.e., causing large obliquity), could simultaneously align the planetary orbits with respect to the binary orbit. Future observations (e.g., GAIA and its synergy with radial velocity and transit observations), which can resolve  3-D orbital motion, will test the orbital {\xxie (anti-)}alignment effect identified in this work.

{\xxie The orbital (anti-)alignment effect is symmetrical with respect to $i_B=90^\circ$ (Fig 4). Systems with initial $i_B>90^\circ$ tend to end up with anti-aligned orbital configurations. 
Recent studies have revealed that some planets are probably on coplanar retrograde orbits
in binary systems, e.g., HD 59686 \citep{Tri18}. 
The mechanism studied in this paper provides a possible channel to understand the formation of such interesting planetary systems. 
Nevertheless, HD 59686 is a close binary system ($a_B=13.6$ AU) with a giant gaseous planet \citep{Ort16}, while this paper focuses on terrestrial planets.
Whether and how the mechanism can apply to such a system is a problem that deserves a future study with specific consideration.  
}

Besides the orbital {\xxie (anti-)}alignment effect, we also find that a binary stellar environment is not always negative to planet formation. {\Xie Both observations \citep{Wan14, Kra16} and simulations \citep{Qui07} have shown that close binaries (i.e., $a_B<20$ AU) may suppress planet formation.  Nevertheless the situation could be reversed in binaries with intermediate separations, i.e., $a_B\sim$ 40$-$200 AU. Indeed, we find that planets can grow up to be significantly more massive (though fewer in numbers) in such intermediate separation binaries as compared to those formed in single star systems.  The on-going TESS mission will detect thousands of planets which are close enough to measure their masses via radial velocity observation, thus providing an opportunity to test this positive binary effect.}

\acknowledgments
We acknowledges support from the National Natural Science Foundation of China (Grant No. 11333002, 11403012 and 11661161014). J.-W. X. is also supported by the Foundation for the Author of National Excellent Doctoral Dissertation (FANEDD) of PR China.


\appendix
\section{Results with Bulirsch-Stoer (BS) algorithm}
We performed more simulations using the Bulirsch-Stoer (BS) algorithm in MERCURY.
These additional simulations confirm our main result, i.e., the orbital alignment effect in intermediate separated binaries and provide more insights into this effect 

\begin{figure}
\includegraphics[width=\textwidth, height=0.5\textwidth]{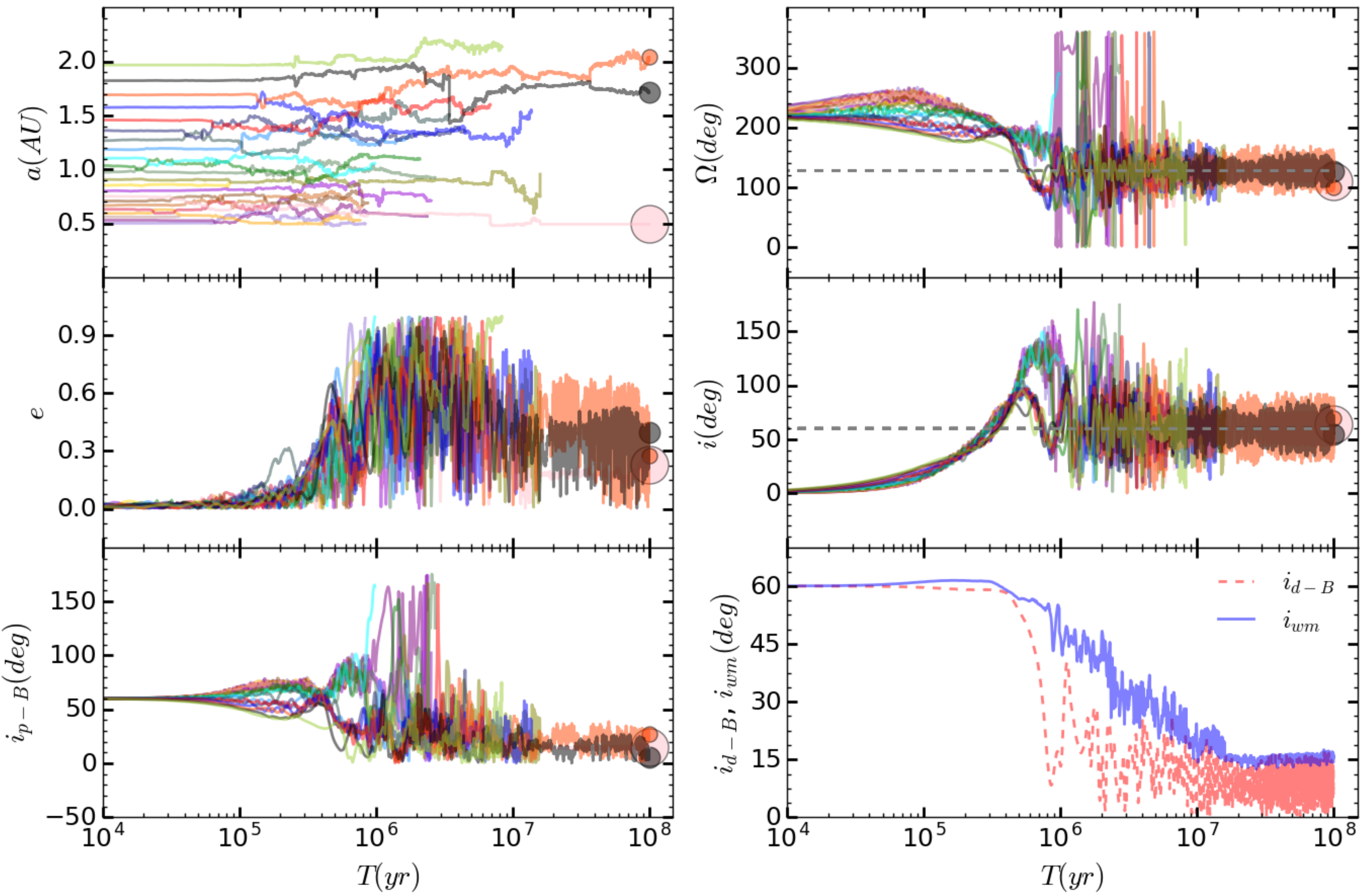}
\caption{Similar to Figure \ref{fig:track} but using the Bulirsch-Stoer (BS) algorithm instead of the Wide-Binary (WB) algorithm. \label{fig:a1}}
\end{figure}

(1) {\it  Effect of integration algorithm.} We performed the same simulation as the standard case in section 3.1, but replaced the Wide-Binary (WB) algorithm with the Bulirsch-Stoer (BS) algorithm. 
The result is plotted in Figure \ref{fig:a1}. 
In this case, there are three embryos left at the end of simulation, i.e., $T=10^8$ yr, with mean inclination $i_{wm}\sim15^\circ$, which is similar to the state of the standard case at $T\sim6\times10^7$ yr in Figure \ref{fig:track}. 
Nevertheless, in the standard case, two of the embryos collided afterwards and the system was further damped to $i_{wm}\sim5^\circ$. 
From this comparison, we see that both the simulations essentially gave the same orbital alignment effect. 
The difference is in final degree of such an alignment, which is likely due to the  simulation time limit, i.e., $T=10^8$ yr. 
It is possible that the system in the case here  (Fig.\ref{fig:a1}) will have one more collision and evolve to a more aligned state with a smaller $i_{wm}$ value after $T=10^8$ yr.

\begin{figure}
\begin{center}
\includegraphics[width=0.7\textwidth, height=0.5\textwidth]{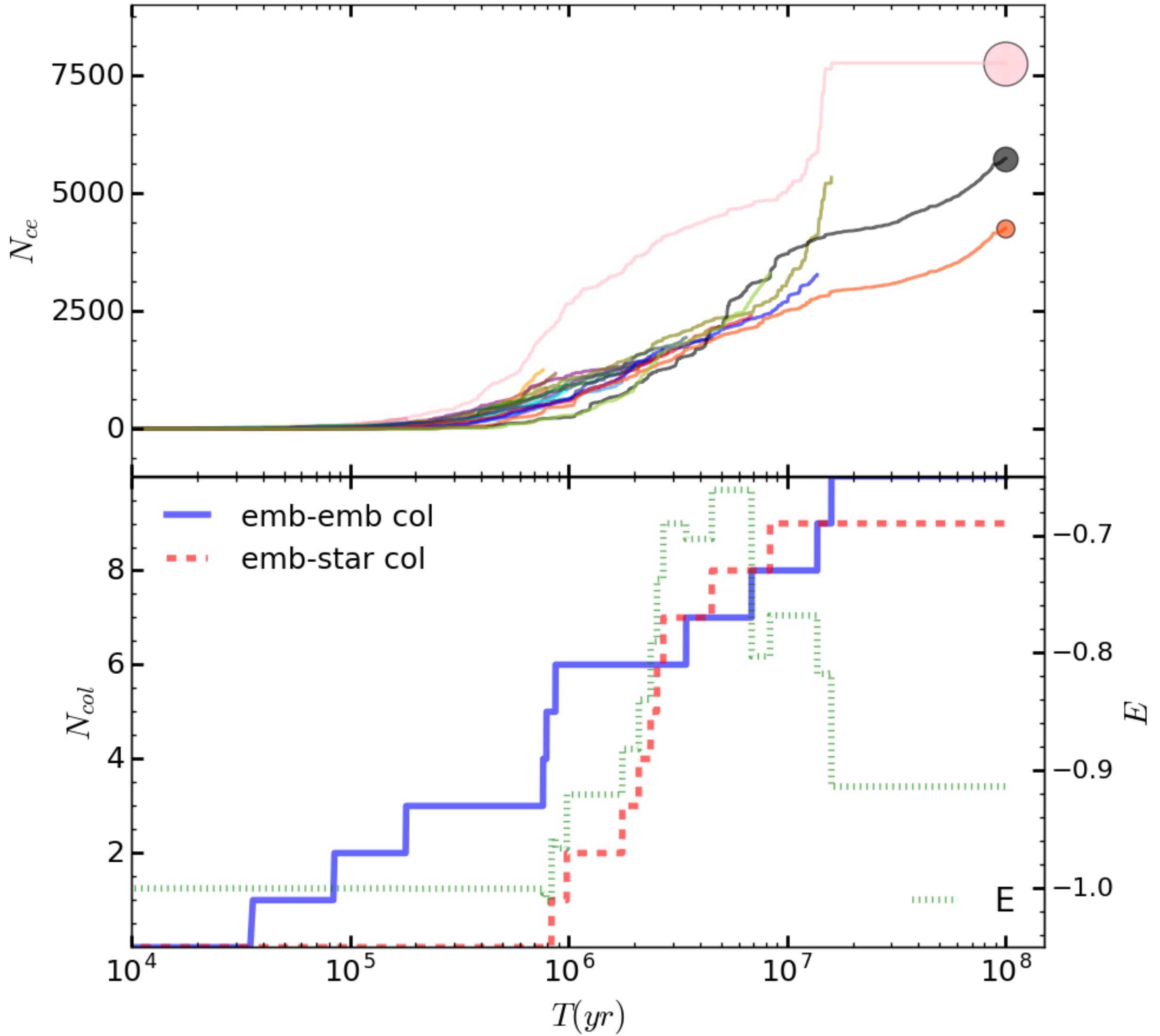}
\caption{Number of close encounters ($N_{ce}$),  number of collisions ($N_{col}$) and total orbital energy of embryos ($E$, normalized by the absolute value of the initial energy) as a function of time in the simulation of Figure \ref{fig:a1}. Note that the energy change before $\sim 10^6$ is too small to be visible in the figure. \label{fig:a2}}
\end{center}
\end{figure}

In this case, we also plot in Figure \ref{fig:a2} the number of close encounters ($N_{ce}$), number of collisions ($N_{col}$) and the total orbital energy of embryos ($E$) as a function of time. We see that most collisions and close encounters occurred during $T\sim10^6-10^7$ yr with significant changes in the total orbital energy, which is in line with the $i_{wm}$ damping as shown in Figure \ref{fig:a1}.  Note, there were two kinds of collisions, i.e., embryo-embryo collision and embryo-star collision. The former reduced the total orbital energy and the latter increased the total orbital energy. There were several early collisions among embryos with similar orbits  (before $T=10^6$ yr when embryos' orbit had not been severely excited) and thus caused slight change in the energy (not visible in Fig. \ref{fig:a2}).

\begin{figure}
\includegraphics[width=\textwidth, height=0.5\textwidth]{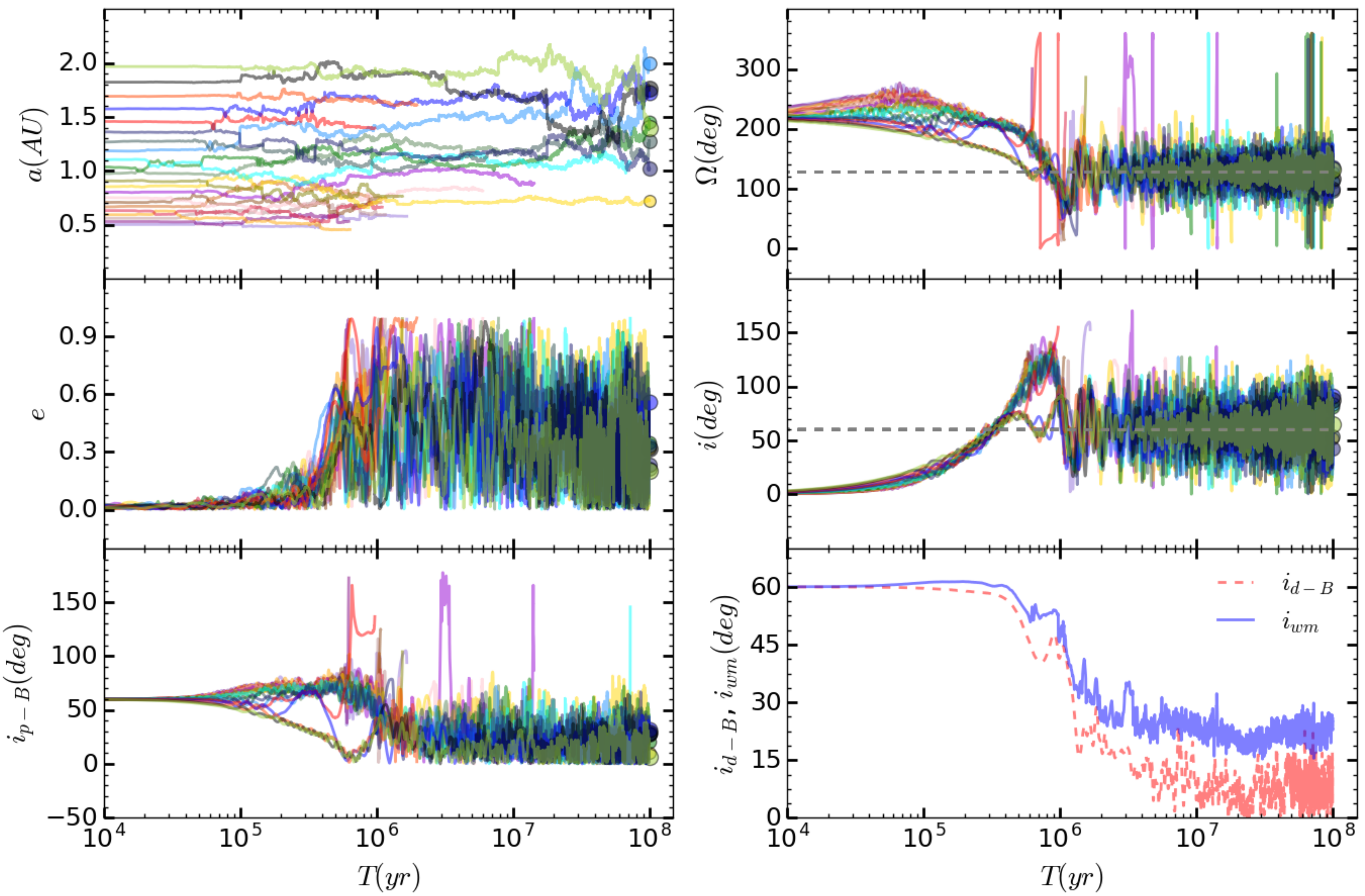}
\caption{Similar to Figure \ref{fig:a1} but turning off embryos' mutual collision by set setting the radii of all embryos as close to zero (note, the embryos can still interact with each other via mutual gravity and may collide with the central star in this case).\label{fig:a3}}
\end{figure}

(2) {\it  Effect of collision.} We performed the same simulation as above but turned off embryos' mutual collisions by artificially setting their radii close to zero.   The results are plotted in Figure \ref{fig:a3}. As can be seen, turning off collision slightly reduced the degree of orbital damping. At the end of simulation, there were  8 embryos left with $i_{wm}\sim22^\circ$ a bit larger than that in the Figure \ref{fig:a1}.  This indicates the major damping source is not collision but dynamical friction via self-gravity. In addition, the alignment between the whole disk and the binary ($i_{d-B}\sim0$) was not affected by collision. 

\begin{figure}
\includegraphics[width=\textwidth, height=0.5\textwidth]{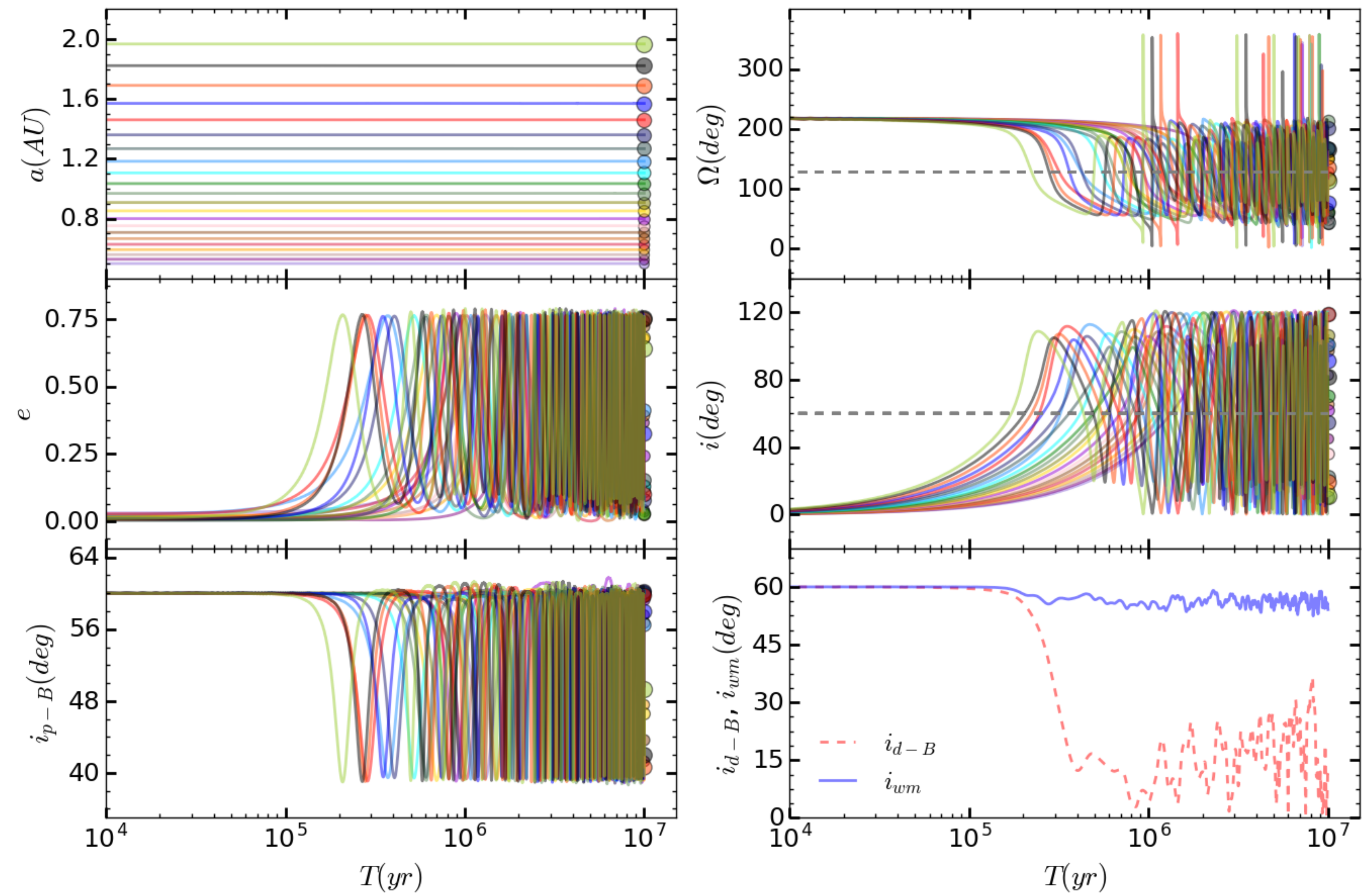}
\caption{Similar to Figure \ref{fig:a1} but turning off embryos' mutual collision and their self-gravity by set setting them as test particles with zero radii and zero masses.\label{fig:a4}}
\end{figure}

(3) {\it  Effect of self-gravity.} We further turned off embryos' self-gravity by setting them as test particles. The results are shown in Figure \ref{fig:a4}. The systems were dominated by binary secular perturbations. The Kozai cycle can be clearly seen from the evolutions of  $e$ and $i_{p-B}$. The orbital differential precession can be clearly seen from the evolution of $\Omega$ and $i$, which caused alignment between the embryo disk and the binary orbit, i.e., $i_{d-B}\sim0$ in less than $10^6$ yr. However, the disk was still very ``thick" or dynamically hot with $i_{wm}=60^\circ$, since all the damping sources (collision and dynamical friction) were not taken into account.\\

Based on above simulations, we learn that the orbital alignment effect requires two crucial processes.  One is a dynamically pumping process with orbital differential precession, which reduces $i_{d-B}$. The other is a dynamically damping process due to dynamical friction and/or collision, which reduces $i_{wm}$.\\

\begin{figure}
\includegraphics[width=\textwidth, height=0.5\textwidth]{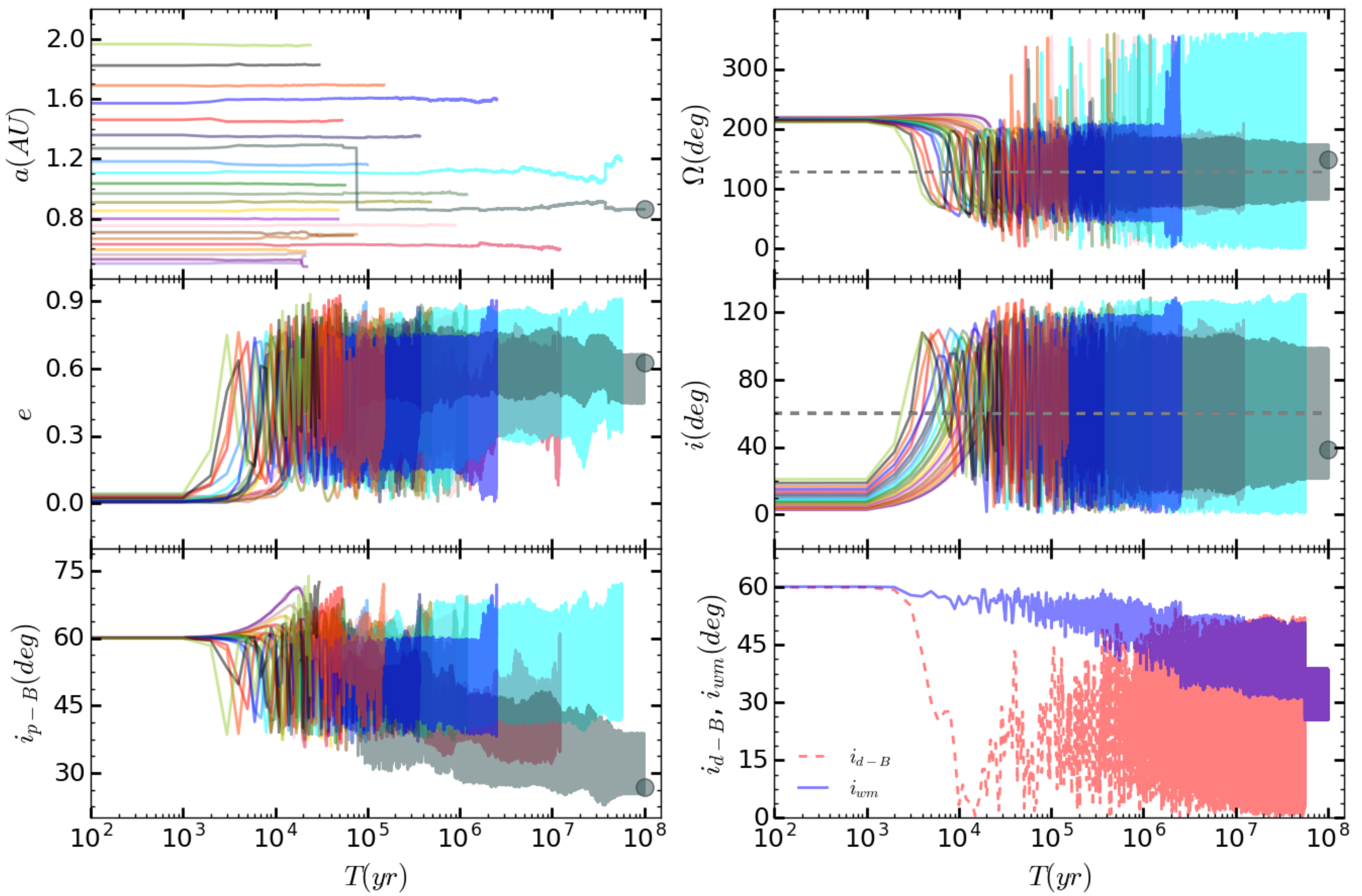}
\caption{Similar to Figure \ref{fig:a1} but setting $a_B=20$ AU instead of $a_B=80$ AU. \label{fig:a5}}
\end{figure}

(4) {\it  Effect of binary separation.}We first consider a much closer binary. 
We performed the same simulation as in Figure \ref{fig:a1}, but set $a_B=20$ AU. 
The results are plotted in Figure \ref{fig:a5}. 
As can be seen,  $i_{d-B}$ reduced toward zero within $\sim10^4$ yr and then fluctuate afterwards. 
The quick reduction of  $i_{d-B}$ is due to the shorter timescale of binary perturbation with a closer binary separation.  
As most embryos were quickly lost  (collided with the primary star) due to strong binary perturbation, the damping became inefficient as can be seen from the slight decrease in $i_{wm}$. 
Finally,  $i_{d-B}$ and $i_{wm}$ converge because only one embryo left in the system.

\begin{figure}
\includegraphics[width=\textwidth, height=0.5\textwidth]{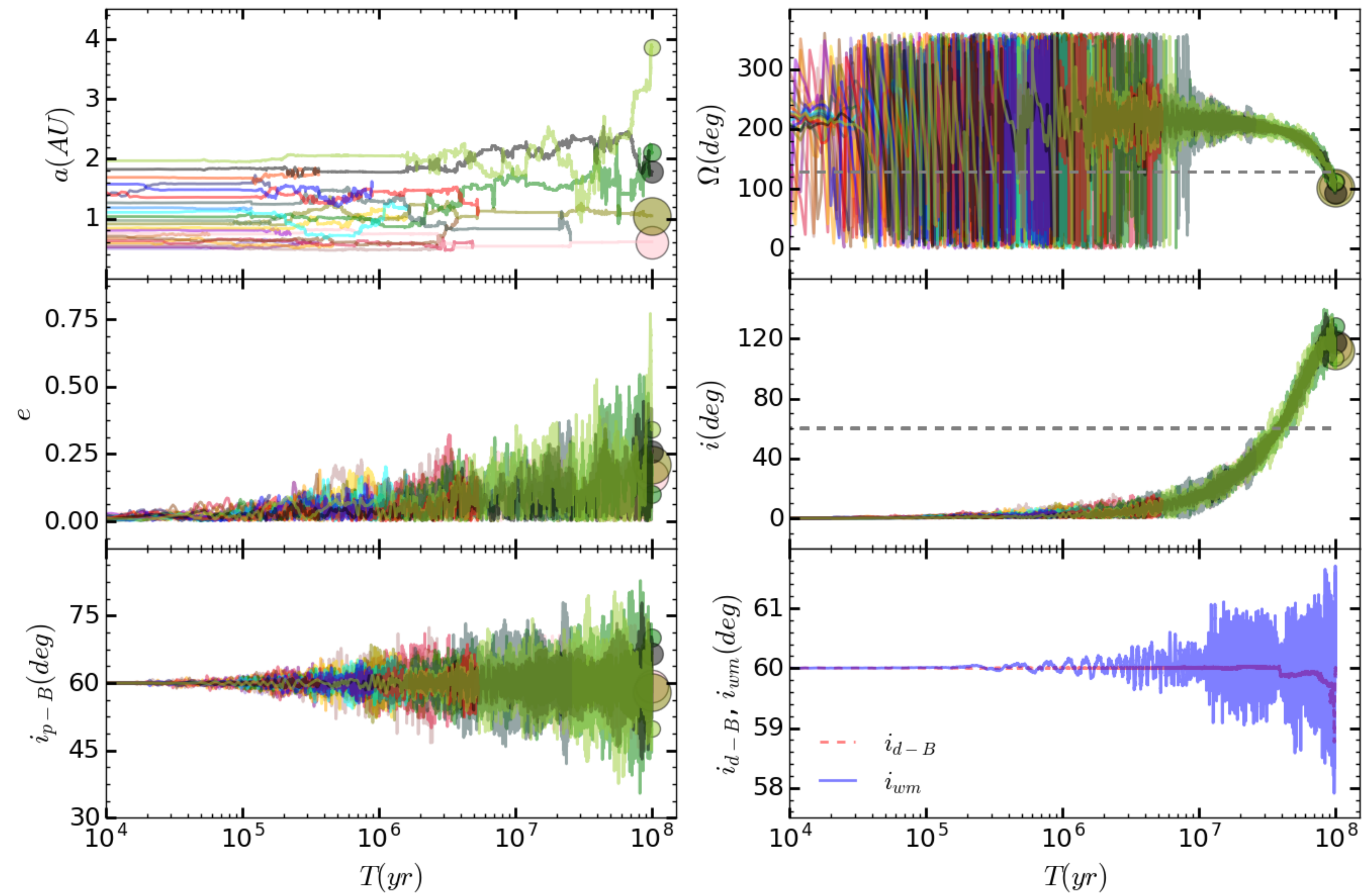}
\caption{Similar to Figure \ref{fig:a1} but setting $a_B=400$ AU instead of $a_B=80$ AU. \label{fig:a6}}
\end{figure}

We then consider a much wider binary. 
We performed the same simulation as in Figure \ref{fig:a1}, but set $a_B=400$ AU. 
The results are plotted in Figure \ref{fig:a6}. 
As can be seen, the dynamical pumping was much weaker due to the wider binary separation. And thus orbital differential precession was largely quenched by embryos' self-gravity. Embryos evolved like a rigid disk as can be seen from their evolutions of $\Omega$ and $i$, and there was no orbital alignment effect; $i_{d-B}$ still maintained large values. 

Based on above simulations, we learn that the orbital alignment effect is most significant in cases of intermediate binary separation (as shown in Figure \ref{fig:depend}) because both orbital differential precession and orbital damping are efficient in such cases.

\clearpage

\clearpage
\startlongtable
\begin{deluxetable}{c|cccccccc}
\tablecaption{Statistics\tablenotemark{1}  at the End of Simulations ($T=10^8$ yr). \label{tab:A}}
\tablehead{
\colhead{Case\tablenotemark{2} } &\colhead{$i_{p-B}$\tablenotemark{3} }& \colhead{$M_r$\tablenotemark{4} } &\colhead{$N_p$\tablenotemark{5} }&\colhead{$M_{max}$\tablenotemark{6} }& \colhead{$e$\tablenotemark{7} } &\colhead{$i$\tablenotemark{8} }  &\colhead{$i_{mul}$\tablenotemark{9} }&\colhead{$L_z$\tablenotemark{10} } }
\startdata
$i0-a20  $ & 4.49&0.93&3.2&1.64&0.12&4.49&6.42&0.96\\
$i0-a40  $ & 7.04&1.00&3.6&1.87&0.18&7.04&8.68&0.98\\
$i0-a60  $ & 5.54&1.00&4.2&1.46&0.15&5.54&8.41&1.00\\
$i0-a80  $ & 6.50&1.00&4.4&1.64&0.17&6.50&9.86&0.99\\
$i0-a100 $  & 8.30&1.00&4.8&1.64&0.16&8.30&12.69&1.00\\
$i0-a150 $  & 9.53&1.00&4.2&1.69&0.17&9.53&14.21&1.00\\
$i0-a200 $  & 7.13&0.99&4.2&1.80&0.13&7.13&11.51&1.00\\
$i0-a400 $  & 4.94&0.99&4.1&1.46&0.13&4.94&8.34&1.00\\
$i0-a600 $  & 7.78&1.00&4.8&1.58&0.14&7.78&12.38&1.00\\
$i0-a800 $  & 7.26&1.00&4.4&1.66&0.16&7.26&11.77&1.00\\
$i0-a1000$ & 6.26&1.00&4.7&1.69&0.14&6.26&10.32&1.00\\
$i30-a20  $& 14.87&0.86&4.8&1.46&0.26&35.55&22.55&0.98\\
$i30-a40  $& 12.41&0.86&4.7&1.57&0.24&30.97&18.18&1.07\\
$i30-a60  $& 11.82&0.82&3.6&1.83&0.27&32.67&19.03&1.10\\
$i30-a80  $& 12.25&0.89&3.9&2.01&0.28&33.98&16.50&1.05\\
$i30-a100 $ & 15.57&0.90&3.0&2.18&0.21&34.29&20.82&1.03\\
$i30-a150 $ & 24.25&0.93&3.2&1.83&0.22&39.67&22.54&1.03\\
$i30-a200 $ & 26.14&0.96&4.1&1.78&0.27&44.72&21.37&1.03\\
$i30-a400 $ & 29.65&1.00&4.4&1.63&0.14&12.31&10.15&1.00\\
$i30-a600 $ & 30.48&1.00&4.5&1.50&0.13&45.76&9.97&1.00\\
$i30-a800 $ & 29.44&1.00&4.5&1.37&0.12&20.47&8.52&1.00\\
$i30-a1000$ & 30.39&1.00&4.5&1.42&0.12&12.40&10.53&1.00\\
$i45-a20 $& 17.77&0.42&2.3&1.10&0.22&45.67&20.39&1.08\\
$i45-a40 $& 16.99&0.59&3.9&1.10&0.31&48.13&25.52&1.32\\
$i45-a60 $& 13.83&0.59&3.5&1.15&0.26&45.63&21.89&1.40\\
$i45-a80 $& 12.73&0.68&2.3&1.92&0.28&45.48&18.34&1.19\\
$i45-a100 $& 14.71&0.66&2.4&1.87&0.27&44.87&22.13&1.22\\
$i45-a150 $& 25.11&0.80&2.2&2.32&0.24&60.36&25.71&1.13\\
$i45-a200 $& 31.93&0.89&2.7&1.92&0.33&44.03&30.66&1.07\\
$i45-a400 $& 44.47&1.00&4.5&1.61&0.17&45.52&13.91&1.00\\
$i45-a600 $& 44.64&1.00&4.8&1.41&0.19&57.19&12.77&1.00\\
$i45-a800 $& 45.50&1.00&4.8&1.51&0.14&24.24&10.00&1.00\\
$i45-a1000$ & 45.28&1.00&5.2&1.53&0.16&15.31&13.08&1.00\\
$i60-a20 $& 39.66&0.10&1.2&0.32&0.43&73.67&18.19&1.07\\
$i60-a40 $& 20.41&0.28&2.7&0.54&0.29&57.30&29.31&1.77\\
$i60-a60 $& 23.41&0.38&3.4&0.71&0.27&60.16&36.78&1.97\\
$i60-a80 $& 15.72&0.37&2.8&0.72&0.25&62.43&22.59&2.04\\
$i60-a100 $& 17.99&0.39&2.3&1.07&0.27&64.54&24.02&1.88\\
$i60-a150 $& 26.08&0.48&1.3&1.61&0.30&56.73&40.53&1.65\\
$i60-a200 $& 34.54&0.66&1.7&1.99&0.31&69.75&33.31&1.42\\
$i60-a400 $& 59.59&1.00&4.2&1.51&0.19&112.10&13.50&1.00\\
$i60-a600 $& 60.29&1.00&5.0&1.55&0.18&54.51&12.26&1.00\\
$i60-a800 $& 60.58&1.00&4.4&1.49&0.13&21.33&9.60&1.00\\
$i60-a1000$ & 60.09&0.99&4.8&1.68&0.15&14.13&13.77&1.00\\
$i80-a20 $& NA &0.00&0.0&NA&NA&NA&NA&NA\\
$i80-a40 $& 64.70&0.07&1.0&0.24&0.42&83.92&NA&2.69\\
$i80-a60 $& 50.35&0.08&1.1&0.25&0.58&83.54&113.38&3.38\\
$i80-a80 $& 43.64&0.08&1.0&0.28&0.48&82.23&NA&3.76\\
$i80-a100 $& 42.76&0.15&1.0&0.53&0.62&81.88&NA&3.39\\
$i80-a150 $& 48.04&0.27&1.0&0.98&0.61&66.09&NA&3.23\\
$i80-a200 $& 35.74&0.21&1.0&0.74&0.60&88.38&NA&4.82\\
$i80-a400 $& 80.35&1.00&4.5&1.66&0.18&68.67&13.85&1.00\\
$i80-a600 $& 79.98&1.00&5.0&1.41&0.12&20.22&9.60&1.00\\
$i80-a800 $& 79.90&0.99&5.2&1.51&0.18&11.69&12.60&1.00\\
$i80-a1000$ & 78.76&1.00&4.4&1.67&0.15&8.80&12.54&1.00\\
$i100-a20 $& NA&0.00&0.0&NA&NA&NA&NA&NA\\
$i100-a40 $& 118.02&0.06&1.0&0.20&0.48&81.16&NA&2.30\\
$i100-a60 $& 130.70&0.08&1.0&0.27&0.50&88.81&NA&3.81\\
$i100-a80 $& 141.12&0.09&1.0&0.31&0.52&92.79&NA&3.42\\
$i100-a100$ & 132.46&0.14&1.0&0.49&0.49&59.74&NA&3.38\\
$i100-a150$ & 129.50&0.26&1.1&0.89&0.61&75.88&10.60&3.11\\
$i100-a200$ & 122.31&0.40&1.0&1.43&0.56&109.82&NA&2.65\\
$i100-a400$ & 98.77&0.99&4.5&1.59&0.22&74.46&17.49&1.00\\
$i100-a600$ & 100.01&1.00&4.5&1.70&0.19&24.39&16.78&1.00\\
$i100-a800$ & 99.50&1.00&4.7&1.62&0.14&11.82&12.27&1.00\\
$i100-a1000$ & 99.86&1.00&4.8&1.46&0.15&8.96&11.82&1.00\\
$i120-a20 $& 140.71&0.09&1.2&0.28&0.49&47.05&47.60&1.00\\
$i120-a40 $& 154.60&0.28&2.6&0.59&0.33&61.69&37.31&1.74\\
$i120-a60 $& 157.34&0.38&3.9&0.60&0.32&63.03&27.05&2.00\\
$i120-a80 $& 159.48&0.36&3.1&0.73&0.35&57.57&25.09&1.96\\
$i120-a100 $& 161.04&0.46&1.7&1.38&0.23&62.64&25.03&1.64\\
$i120-a150 $& 150.01&0.54&1.3&1.82&0.25&66.18&30.98&1.59\\
$i120-a200 $& 138.50&0.73&1.9&2.09&0.44&42.60&31.00&1.25\\
$i120-a400 $& 119.87&1.00&4.5&1.53&0.16&115.21&11.59&1.00\\
$i120-a600 $& 118.95&1.00&4.5&1.54&0.15&51.13&12.15&1.00\\
$i120-a800 $& 119.17&1.00&4.5&1.66&0.14&23.46&12.16&1.00\\
$i120-a1000$ & 120.01&1.00&4.0&1.76&0.13&13.28&11.66&1.00\\
$i135-a20 $& 156.33&0.46&2.7&1.08&0.23&51.31&37.95&1.06\\
$i135-a40 $& 162.29&0.63&3.3&1.38&0.27&45.47&27.37&1.28\\
$i135-a60 $& 164.77&0.61&3.8&1.16&0.30&45.40&21.41&1.38\\
$i135-a80 $& 160.45&0.64&2.8&1.58&0.31&46.62&27.70&1.25\\
$i135-a100 $& 165.24&0.63&2.1&1.80&0.23&45.19&24.52&1.25\\
$i135-a150 $& 159.25&0.78&2.2&2.13&0.25&49.84&22.59&1.18\\
$i135-a200 $& 149.76&0.86&2.5&2.10&0.28&47.52&21.50&1.10\\
$i135-a400 $& 134.89&1.00&4.1&1.76&0.14&46.72&8.76&1.00\\
$i135-a600 $& 134.63&1.00&5.3&1.45&0.16&59.79&13.46&1.00\\
$i135-a800 $& 134.83&0.99&3.8&1.80&0.13&25.70&10.53&1.01\\
$i135-a1000$ & 134.22&1.00&4.5&1.47&0.11&13.63&9.53&1.00\\
$i150-a20 $& 164.96&0.86&4.6&1.58&0.20&33.36&21.17&1.00\\
$i150-a40 $& 170.34&0.87&4.0&1.63&0.23&30.13&13.41&1.07\\
$i150-a60 $& 167.56&0.83&3.3&2.04&0.23&31.42&18.76&1.08\\
$i150-a80 $& 165.97&0.88&3.6&2.05&0.21&33.46&21.59&1.06\\
$i150-a100 $& 162.77&0.91&3.7&2.27&0.26&35.51&24.30&1.05\\
$i150-a150 $& 159.10&0.95&3.8&2.24&0.35&34.05&25.95&1.01\\
$i150-a200 $& 157.30&0.98&3.8&2.02&0.33&39.03&28.89&1.01\\
$i150-a400 $& 149.88&1.00&4.2&1.39&0.11&16.59&9.21&1.00\\
$i150-a600 $& 149.28&1.00&4.4&1.49&0.17&49.32&11.90&1.00\\
$i150-a800 $& 148.73&1.00&4.5&1.61&0.13&23.40&14.30&1.00\\
$i150-a1000 $& 147.52&1.00&5.3&1.57&0.14&14.15&11.81&1.00\\
single & NA&1.00&5.9&1.38&0.17&8.18&13.03& NA\\
\enddata 
\tablenotetext{1}{These values are average on 10 realizations}
\tablenotetext{2}{Code, $iX-aY$, indicates the simulation case with initial binary inclination $i_B=X$ deg, and binary separation $a_B=Y$ AU. }
\tablenotetext{3}{The average inclination relative to the binary orbital plane (in deg).}
\tablenotetext{4}{The fraction of the remaining planetary mass.}
\tablenotetext{5}{The number of remaining planets/embryos.}
\tablenotetext{6}{The mass of the largest planet (in $M_{\oplus}$).}
\tablenotetext{7}{The average eccentricity.}
\tablenotetext{8}{The average inclination with respect to the initial proto-planetary disk plane (in deg).}
\tablenotetext{9}{The average mutual orbital inclination of remaining planets (in deg).}
\tablenotetext{10}{The Z-component (perpendicular to the binary orbital plane) of angular momentum per unit mass (normalized by the initial value $L_{z0}$).\\}

\end{deluxetable}


\begin{thebibliography}{}
\bibitem[Artymowicz \& Lubow(1994)]{AL94} Artymowicz, P., \& Lubow, S.~H.\ 1994, \apj, 421, 651 
\bibitem[Bate et al.(2000)]{Bat00} Bate, M.~R., Bonnell, I.~A., Clarke, C.~J., et al.\ 2000, \mnras, 317, 773 

\bibitem[Batygin(2012)]{Bat12} Batygin, K.\ 2012, \nat, 491, 418
\bibitem[Brinch et al.(2016)]{Bri16} Brinch, C., J{\o}rgensen, J.~K., Hogerheijde, M.~R., Nelson, R.~P., \& Gressel, O.\ 2016, \apjl, 830, L16

\bibitem[Chambers (1999)]{Cha99} Chambers J. E. \ 1999, MNRAS, 304, 793
\bibitem[Chambers et al. (2002)]{Cha02} Chambers, J. E., Quintana, E. V., Duncan, M. J., \& Lissauer, J. J. \ 2002, AJ, 123, 2884
\bibitem[Desidera \& Barbieri(2007)]{DB07} Desidera, S., \& Barbieri, M.\ 2007, \aap, 462, 345
\bibitem[Doyle et al.(2011)]{Doy11} Doyle, L.~R., Carter, J.~A., Fabrycky, D.~C., et al.\ 2011, Science, 333, 1602 
\bibitem[Dumusque et al.(2012)]{Dum12} Dumusque, X., Pepe, F., Lovis, C., et al.\ 2012, \nat, 491, 207 

\bibitem[Duquennoy \& Mayor(1991)]{DM91} Duquennoy, A., \& Mayor, M.\ 1991, \aap, 248, 485
\bibitem[Fabrycky \& Tremaine(2007)]{FT07} Fabrycky, D., \& Tremaine, S.\ 2007, \apj, 669, 1298
\bibitem[Fern{\'a}ndez-L{\'o}pez et al.(2017)]{Fer17} Fern{\'a}ndez-L{\'o}pez, M., Zapata, L.~A., \& Gabbasov, R.\ 2017, \apj, 845, 10

\bibitem[Fragner \& Nelson(2010)]{FN10} Fragner, M.~M., \& Nelson, R.~P.\ 2010, \aap, 511, A77 
\bibitem[Fragner et al.(2011)]{Fra11} Fragner, M.~M., Nelson, R.~P., \& Kley, W.\ 2011, \aap, 528, A40 
\bibitem[Fu et al.(2015)]{Fu15} Fu, W., Lubow, S.~H., \& Martin, R.~G.\ 2015, \apj, 807, 75 
\bibitem[Haghighipour(2006)]{Hag06} Haghighipour, N.\ 2006, \apj, 644, 543 
\bibitem[Haghighipour \& Raymond(2007)]{HR07} Haghighipour, N., \& Raymond, S.~N.\ 2007, \apj, 666, 436 

\bibitem[Haghighipour(2010)]{Hag10} Haghighipour, N.\ 2010, Planets in Binary Star Systems, 366,  
\bibitem[Hale(1994)]{Hal94} Hale, A.\ 1994, \aj, 107, 306 

\bibitem[Hansen \& Murray (2013)]{HM13} Hansen, B. M. S., \& Murray, N. \ 2013, \apj, 775, 53
\bibitem[Hatzes et al.(2003)]{Hat03} Hatzes, A.~P., Cochran, W.~D., Endl, M., et al.\ 2003, \apj, 599, 1383 

\bibitem[Howard et al.(2012)]{How12} Howard, A.~W., Marcy, G.~W., Bryson, S.~T., et al.\ 2012, \apjs, 201, 15 
\bibitem[Jensen et al.(2004)]{Jen04} Jensen, E.~L.~N., Mathieu, R.~D., Donar, A.~X., \& Dullighan, A.\ 2004, \apj, 600, 789 
\bibitem[Jensen \& Akeson(2014)]{JA14} Jensen, E.~L.~N., \& Akeson, R.\ 2014, \nat, 511, 567 
\bibitem[Kley \& Nelson(2008)]{KN08} Kley, W., \& Nelson, R.~P.\ 2008, \aap, 486, 617 

\bibitem[Kokubo et al. (2006)]{Kok06} Kokubo, E., Kominami, J., \& Ida, S. \ 2006, \apj, 642, 1131
\bibitem[Kraus et al.(2016)]{Kra16} Kraus, A.~L., Ireland, M.~J., Huber, D., Mann, A.~W., \& Dupuy, T.~J.\ 2016, \aj, 152, 8 
\bibitem[Larwood et al.(1996)]{Lar96} Larwood, J.~D., Nelson, R.~P., Papaloizou, J.~C.~B., \& Terquem, C.\ 1996, \mnras, 282, 597 
\bibitem[Lee et al.(2017)]{Lee17} Lee, J.-E., Lee, S., Dunham, M.~M., et al.\ 2017, Nature Astronomy, 1, 0172
\bibitem[Lubow \& Ogilvie(2000)]{LO00} Lubow, S.~H., \& Ogilvie, G.~I.\ 2000, \apj, 538, 326 
\bibitem[Martin et al.(2014)]{Mar14} Martin, R.~G., Nixon, C., Lubow, S.~H., et al.\ 2014, \apjl, 792, L33
\bibitem[Marzari et al.(2009)]{Mar09} Marzari, F., Th{\'e}bault, P., \& Scholl, H.\ 2009, \aap, 507, 505 
\bibitem[Mayor et al.(2011)]{May11} Mayor, M., Marmier, M., Lovis, C., et al.\ 2011, arXiv:1109.2497 
\bibitem[Monin et al.(2006)]{Mon06} Monin, J.-L., M{\'e}nard, F., \& Peretto, N.\ 2006, \aap, 446, 201 
\bibitem[Monin et al.(2007)]{Mon07} Monin, J.-L., Clarke, C.~J., Prato, L., \& McCabe, C.\ 2007, Protostars and Planets V, 395 

\bibitem[O'Brien et al.(2006)]{Obr06} O'Brien, D.~P., Morbidelli, A., \& Levison, H.~F.\ 2006, \icarus, 184, 39 

\bibitem[Ortiz et al.(2016)]{Ort16} Ortiz, M., Reffert, S., Trifonov, T., et al.\ 2016, \aap, 595, A55

\bibitem[Quintana et al. (2002)]{Qui02} Quintana, E. V., Lissauer, J. J., Chambers, J. E., \& Duncan, M. J. \ 2002, \apj, 576, 982

\bibitem[Quintana et al.(2007)]{Qui07} Quintana, E.~V., Adams, F.~C., Lissauer, J.~J., \& Chambers, J.~E.\ 2007, \apj, 660, 807 

\bibitem[Rafikov \& Silsbee(2015)]{RS15a} Rafikov, R.~R., \& Silsbee, K.\ 2015, \apj, 798, 69 
\bibitem[Rafikov \& Silsbee(2015)]{RS15b} Rafikov, R.~R., \& Silsbee, K.\ 2015, \apj, 798, 70 

\bibitem[Raghavan et al.(2010)]{Rag10} Raghavan, D., McAlister, H.~A., Henry, T.~J., et al.\ 2010, \apjs, 190, 1 

\bibitem[Takeda et al.(2008)]{Tak08} Takeda, G., Kita, R., \& Rasio, F.~A.\ 2008, \apj, 683, 1063-1075

\bibitem[Th{\'e}bault et al.(2006)]{The06} Th{\'e}bault, P., Marzari, F., \& Scholl, H.\ 2006, \icarus, 183, 193 
\bibitem[Th{\'e}bault et al.(2008)]{The08} Th{\'e}bault, P., Marzari, F., \& Scholl, H.\ 2008, \mnras, 388, 1528 
\bibitem[Th{\'e}bault et al.(2009)]{The09} Th{\'e}bault, P., Marzari, F., \& Scholl, H.\ 2009, \mnras, 393, L21 
\bibitem[Thebault(2011)]{The11} Thebault, P.\ 2011, Celestial Mechanics and Dynamical Astronomy, 111, 29 

\bibitem[Thebault \& Haghighipour(2014)]{TH14} Thebault, P., \& Haghighipour, N.\ 2014, arXiv:1406.1357 
\bibitem[Tremaine \& Davis(2014)]{TD14} Tremaine, S., \& Davis, S.~W.\ 2014, \mnras, 441, 1408 

\bibitem[Trifonov et al.(2018)]{Tri18} Trifonov, T., Lee, M.~H., Reffert, S., \& Quirrenbach, A.\ 2018, \aj, 155, 174

\bibitem[Wang et al.(2014)]{Wan14} Wang, J., Xie, J.-W., Barclay, T., \& Fischer, D.~A.\ 2014, \apj, 783, 4 
\bibitem[Williams et al.(2014)]{Wil14} Williams, J.~P., Mann, R.~K., Di Francesco, J., et al.\ 2014, \apj, 796, 120
\bibitem[Wu \& Murray(2003)]{WM03} Wu, Y., \& Murray, N.\ 2003, \apj, 589, 605
\bibitem[Xie \& Zhou(2008)]{Xie08} Xie, J.-W., \& Zhou, J.-L.\ 2008, \apj, 686, 570-579 
\bibitem[Xie \& Zhou(2009)]{Xie09} Xie, J.-W., \& Zhou, J.-L.\ 2009, \apj, 698, 2066 

\bibitem[Xie et al.(2010)]{Xie10a} Xie, J.-W., Zhou, J.-L., \& Ge, J.\ 2010, \apj, 708, 1566 
\bibitem[Xie et al.(2010)]{Xie10b} Xie, J.-W., Payne, M.~J., Th{\'e}bault, P., Zhou, J.-L., \& Ge, J.\ 2010, \apj, 724, 1153 
\bibitem[Xie et al.(2011)]{Xie11} Xie, J.-W., Payne, M.~J., Th{\'e}bault, P., Zhou, J.-L., \& Ge, J.\ 2011, \apj, 735, 10 
\bibitem[Zanazzi \& Lai(2017a)]{ZL17a} Zanazzi, J.~J., \& Lai, D.\ 2017, arXiv:
1711.03138 
\bibitem[Zanazzi \& Lai(2017b)]{ZL17b} Zanazzi, J.~J., \& Lai, D.\ 2017, arXiv:1712.07655 
\bibitem[Zsom et al.(2011)]{Zso11} Zsom, A., S{\'a}ndor, Z., \& Dullemond, C.~P.\ 2011, \aap, 527, A10 


\end{thebibliography}
\end{document}